\documentclass[%
 reprint,
 amsmath,amssymb,
 aps,
]{revtex4-1}

\usepackage{graphicx}
\usepackage{dcolumn}
\usepackage{bm}

\renewcommand{\a}{\alpha}
\renewcommand{\b}{\beta}
\newcommand{\g}{\gamma}
\newcommand{\bea}{\begin{eqnarray}}
\newcommand{\eea}{\end{eqnarray}}
\newcommand{\f}[2]{\frac{#1}{#2}}
\newcommand{\eq}{&=&}
\newcommand{\nn}{\nonumber \\ }
\newcommand{\ve}{\varepsilon}
\renewcommand{\d}{\delta}
\newcommand{\area}{\int_{-\infty}^\infty }

\renewcommand{\l}{\lambda}
\newcommand{\p}{\partial}
\newcommand{\pp}[2]{\f{\p #1}{\p #2}}

\newcommand{\sref}[1]{Eq (\ref{#1})}
\newcommand{\citeauthorname}[2]{{#1} {#2}}
\newcommand{\citebook}[4]{{#1} {\it #2} ({#3}, {#4}).}
\newcommand{\citepaper}[4]{{#1} {#3} ({#4}).}
\begin{document}

\preprint{APS/123-QED}

\title{Portfolio Optimization Problem with Non-identical Variances of Asset Returns using Statistical Mechanical Informatics}

\author{Takashi Shinzato}
\email{takashi.shinzato@r.hit-u.ac.jp}
 \affiliation{
Mori Arinori Center for Higher Education and Global Mobility,
Hitotsubashi University, 
Tokyo, 1868601, Japan.}

\date{\today}

\begin{abstract}
The portfolio optimization problem in which the variances of the return rates of assets are not identical 
is analyzed in this paper using the methodology of statistical 
 mechanical informatics, specifically, replica analysis. 
We define two characteristic quantities of an optimal portfolio, namely, minimal investment risk and concentrated investment level, 
in order to solve the portfolio optimization problem
 and analytically determine their asymptotical behaviors using replica 
 analysis. 
Moreover, numerical experiments were performed, and a comparison between the results of our simulation and those obtained via replica analysis validated our proposed method.
\begin{description}
\item[PACS number(s)]
{89.65.Gh}, {89.90.+n}, {02.50.-r}
\end{description}
\end{abstract}
\maketitle


\section{Introduction}
Investment is an economic activity in which one pays 
a cost according to the reward that can be expected in the 
future\cite{Bodie,Luenberger}. 
In 1952, in the field of investment science, 
Markowitz introduced the portfolio optimization problem and constructed as a framework of investment management portfolios
with relatively low risk that do not rely on either experience or intuition\cite{Markowitz1952,Markowitz1959}.
More concretely,
the case of investing a finite budget into several assets appropriately is considered,  
for instance, 
Markowitz proposed the investment behavior policy 
that a portfolio that minimizes the expectation of investment risk, defined as 
the variance of the expected returns under {the sum of expected 
returns being constant}, be regarded as an optimal portfolio and analyzed 
what investment strategy could realize such an optimal portfolio.
Konno et al. proposed the absolute deviation model, in which 
the sum of the absolute value of the expected return over all periods 
is regarded as the investment risk, rather than the sum of the square of 
the expected return, and 
indicated that 
the portfolio of minimal expected investment risk of 
the absolute deviation model is consistent with 
the portfolio of minimal expected investment risk of 
the mean-variance model which was introduced by Markowitz\cite{Konno}. 
Rockafellar et al. introduced 
the expected shortfall model that assesses which portfolios have  
an investment risk which is not below a confidence level, 
taking into account the risk of expected returns\cite{Rockafellar}.

In previous decades, portfolio optimization problems which could not
be easily solved using the analytical approach developed in operations 
research have been solved using statistical 
mechanical informatics\cite{Ciliberti,Pafka,Shinzato-Yasuda2015-BP,Shinzato2015-self-averaging}. 
For instance, 
Ciliberti et al. 
analyzed the typical behavior of 
the minimal investment risks of the absolute deviation model and the expected 
shortfall model which were derived in the limit of absolute zero temperature using replica analysis developed in statistical mechanical informatics\cite{Ciliberti}.
Furthermore, Pafka et al. proposed 
a method for excluding 
noise from the variance-covariance matrix of asset returns 
using random matrix theory which has been improved in econophysics, 
evaluated 
the concentrated investment level using the Mar$\check{\rm c}$enko-Pastur law, 
and compared the estimated returns and the actual returns\cite{Pafka}. 
Although a computation 
proportional to the cube of the number of assets is needed in order to 
produce straightforwardly the inverse of the variance-covariance matrix in the case of the mean-variance model,
Shinzato et al. 
developed a faster algorithm in order to derive the optimal portfolio for the risk minimization problem with a budget constraint
by using belief propagation from statistical mechanical informatics\cite{Shinzato-Yasuda2015-BP}.
Shinzato also indicated that 
the minimal investment risk and its concentrated investment level 
satisfy the self-averaging property by using replica analysis, which supports the theoretical approach of 
Ciliberti et al.\cite{Shinzato2015-self-averaging}.

Such previous works {partly discuss the potential of the optimal 
portfolio in} their investment systems by using 
analytical approaches developed in cross-disciplinary research 
involving the portfolio optimization problem and statistical mechanical informatics. However, 
these works do not reveal their systems' full potential. 
In practice, these works primarily consider the portfolio optimization problem under the assumption that 
the returns of assets are independently and identically distributed or 
that the return of each asset is independently distributed and is not 
identical but that the variances of return rates of the assets are the same, for simplicity. 
This assumption ignores the fact that there exist risk-free assets and  risky assets, such as national government bonds and corporate bonds of small and medium-sized firms, in actual securities market. 
Since 
there exist a wide variety of types of assets, 
previous methods\cite{Ciliberti,Pafka,Shinzato-Yasuda2015-BP,Shinzato2015-self-averaging} are insufficient for 
handling investment management in an actual securities market.

Therefore, in this paper, we analyze 
the portfolio optimization problem, especially, the mean-variance model, 
 in which 
the return rates of assets in the market are independently distributed and 
have different variances,
and assess 
the typical behaviors of the minimal investment risk per asset and its 
concentrated investment level using 
replica analysis. Numerical experiments using a belief propagation 
algorithm\cite{Shinzato-Yasuda2015-BP} were performed 
and a comparison between 
the results of our simulation and 
those obtained via replica analysis validated our proposed method.

\section{Materials and Methods}

\subsection{Mean-variance model}
In this subsection, 
we concretely introduce one of the most representative risk 
management models applied to the portfolio 
optimization problem, the mean-variance model. First, as the securities 
market handled in the present paper, 
we assume that the investor can invest
in $N$ assets without constraints on short selling
and the investment ratio of the portfolio invested in asset $i(=1,\cdots,N)$ is represented as 
$w_i\in{\bf R}$ and the total portfolio 
is expressed as the $N$-dimensional vector $\vec{w}=(w_1,\cdots,w_N)^{\rm 
T}\in{\bf R}^N$, where notation ${\rm T}$ herein means 
the transposition of a vector or matrix.
Furthermore, 
we assume 
here that 
the distribution of the return rates of $N$ assets across $p$ scenarios is known, 
and the return rate of asset $i$ in scenario $\mu(=1,\cdots,p)$
is $x_{i\mu}'$, where the expectation of the return rate of asset $i$ is 
represented as $r_i=E[x_{i\mu}']$, which is known.
From this, 
the return rates of all assets in $p$ scenarios are known at time of investing, 
the investment risk 
 of the mean-variance model ${\cal H}(\vec{w}|X)$
 of portfolio $\vec{w}$ is defined as follows:
\bea
{\cal H}(\vec{w}|X)\eq
\f{1}{2N}\sum_{\mu=1}^p
\left(
\sum_{i=1}^Nx_{i\mu}'w_i
-\sum_{i=1}^Nr_iw_i
\right)^2\nn
\eq\f{1}{2}
\sum_{\mu=1}^p
\left(\f{1}{\sqrt{N}}
\sum_{i=1}^Nx_{i\mu}w_i
\right)^2\label{eq1},
\eea
where 
the normalized asset return rate $x_{i\mu}=x_{i\mu}'-r_i$ has the mean normalized to zero for simplicity of discussion below (hereafter, we use $x_{i\mu}$ not $x_{i\mu}'$ where this does not cause confusion in our discussion). In addition, 
the return rectangular 
matrix $X=\left\{\f{x_{i\mu}}{\sqrt{N}}\right\}\in{\cal M}_{N\times p}$ 
whose components are normalized return rates is used. Next, 
as a budget constraint, the sum of investment ratios  
\bea
\sum_{i=1}^Nw_i\eq N\label{eq2},
\eea
is employed. In this work, 
as in cross-disciplinary research fields 
involving the portfolio optimization problem and statistical mechanical 
informatics, so as to simplify our discussion, 
we include a budget constraint only, that is, we do not include an expected returns constraint. 
Note that 
the budget constraint in previous works in the operations research field
is $\sum_{i=1}^Nw_i=1$, not \sref{eq2}.
We explain our difference in choice below.

The mean-variance model handled in the present paper, then, is 
reformulated 
as an optimization problem in which we minimize the investment risk ${\cal H}(\vec{w}|X)$ given in \sref{eq1}
with respect to portfolio $\vec{w}$ under the budget constraint 
\sref{eq2}. In general, when the rank of the return matrix 
$X=\left\{\f{x_{i\mu}}{\sqrt{N}}\right\}\in{\cal M}_{N\times p}$ is 
$N$, in brief, when  $N<p$, it can be shown that 
the optimal portfolio is uniquely determined as follows:
\bea
\vec{w}\eq\f{NJ^{-1}\vec{e}}{\vec{e}^{\rm T}J^{-1}\vec{e}}\label{eq3}, 
\eea
where the constant vector is 
$\vec{e}=(1,\cdots,1)^{\rm T}\in{\bf R}^N$ and 
the variance-covariance matrix 
$J=\left\{J_{ij}\right\}=XX^{\rm T}\in{\cal M}_{N\times N}$, 
gives 
the covariance between asset $i$ and asset $j$, 
\bea
J_{ij}\eq\f{1}{N}
\sum_{\mu=1}^px_{i\mu}x_{j\mu}.\label{eq4}
\eea 
Note that 
$J^{-1}$, the inverse matrix of variance-covariance matrix $J$, exists when $N<p$, since the ranks of the return matrix $X$ and 
{the} variance-covariance matrix $J$ are assumed to be $N$, and the
optimal portfolio is uniquely determined. 
However, when $N>p$,
since $J$ is not assumed to be invertible,
the optimal portfolio is not uniquely determined. Hereafter, we limit our discussion to the
$N<p$ case.

We will consider the following two statistics
as feature quantities which can characterize the potential of a portfolio:  
\bea
\label{eq5}\ve\eq\f{1}{N}{\cal H}(\vec{w}|X),\\
\label{eq6}q_w\eq\f{1}{N}
\sum_{i=1}^Nw_i^2,
\eea
where $\ve$ defined by \sref{eq5} is the investment risk per asset and 
$q_w$ defined by \sref{eq6} is the concentrated investment level.
The nature of $\ve$ can be intuitively comprehended; however, 
since the properties of 
$q_w$ have not been sufficiently discussed in the literature and $q_w$ is not widely used in operations research, 
the properties of  concentrated investment level will be briefly explained here.
For two typically considered investment strategies, the equipartition investment 
strategy (EIS) and the concentrated investment strategy (CIS), 
the portfolios are represented by $\vec{w}^{\rm EIS}=(1,1,\cdots,1)^{\rm T}\in{\bf 
R}^N$ and $\vec{w}^{\rm CIS}=(N,0,\cdots,0)^{\rm T}\in{\bf 
R}^N$, respectively. In the CIS case, we assume that the total budget is invested in asset $1$
to simply our discussion. 
Then, 
the concentrated investment levels of the investment strategies are 
$q_{w}^{{\rm EIS}}=1$ and $q_{w}^{{\rm CIS}}=N$, respectively, 
and we see that $q_w^{{\rm EIS}}<q_w^{{\rm CIS}}$. That is, 
when the concentrated investment level of an investment strategy is small, 
that investment strategy is regarded as 
the equipartition investment 
strategy, and otherwise the investment strategy is treated as 
the concentrated investment 
strategy. 
From this, it is clear that 
the concentrated investment level 
can be used as a measure of 
the degree of investment diversification.

Three points should be noted at this point. 
First, expanding on the above discussion, 
from the definition of the concentrated investment level in 
\sref{eq6}, $q_w\ge1$ is satisfied. This follows 
from using the mean square of the portfolio, $\f{1}{N}\sum_{i=1}^Nw_i^2$, 
and the mean of the portfolio, $\f{1}{N}\sum_{i=1}^Nw_i$, as follows:
\bea
q_w-1\eq\f{1}{N}\sum_{i=1}^Nw_i^2-\left(\f{1}{N}
\sum_{i=1}^Nw_i
\right)^2\nn
\eq\f{1}{N}
\sum_{i=1}^N
\left(w_i-\f{1}{N}\sum_{i=1}^Nw_i\right)^2\label{eq7}.
\eea
This bound implies that
the concentrated investment level in the equipartition investment 
strategy is a 
lower bound on the possible concentrated investment level, while 
the concentrated investment level
in the concentrated investment strategy is increasing with 
increasing concentration level of investing into 
specific assets. 

Second, 
related to the first point, 
if the definitions of the concentrated investment level 
and 
the budget constraint
follow the formulation widely used in
operations research, for instance, 
$q_w^{(1)}=\sum_{i=1}^N(w_i^{(1)})^2
$ and $\sum_{i=1}^Nw_i^{(1)}=1$, 
then concentrated investment level 
$q_w^{(1)}=\sum_{i=1}^N(w_i^{(1)})^2$ does not easily provide statistical interpretations like that in \sref{eq7}.  In 
addition, under the two budget constraints 
$\sum_{i=1}^Nw_i^{(1)}=1$ and $\sum_{i=1}^Nw_i^{(N)}=N$,
optimal portfolios in \sref{eq1} are denoted as 
$\vec{w}^{(1)}$ and 
$\vec{w}^{(N)}$, respectively, 
and the portfolio ratios of asset $i$ to asset $j$ for the two optimal portfolios are consistent with each other,  
$w_i^{(1)}/w_j^{(1)}=w_i^{(N)}/w_j^{(N)}$, 
since $\vec{w}^{(N)}=N\vec{w}^{(1)}$. 
Because of this agreement,  
we choose to use the budget constraint \sref{eq2}, 
since the concentrated investment level alone
is sufficient for statistical 
interpretations, unlike the 
budget constraint widely used in operations 
research. 

Lastly, we substitute the optimal portfolio given in 
\sref{eq3} into $\ve$ in \sref{eq5} and $q_w$ in
\sref{eq6}, {which gives}
\bea
\label{eq8}\ve\eq\f{1}{2\left(\f{1}{N}\vec{e}^{\rm T}J^{-1}\vec{e}\right)},\\
\label{eq9}q_w\eq\f{\f{1}{N}\vec{e}^{\rm T}J^{-2}\vec{e}}
{\left(\f{1}{N}\vec{e}^{\rm T}J^{-1}\vec{e}\right)^2
}.
\eea
However, 
the computation of the inverse of the variance-covariance matrix is 
$O(N^3)$, 
while the number of assets $N$ is large, and so 
$\f{1}{N}\vec{e}^{\rm T}J^{-1}\vec{e}$ and 
$\f{1}{N}\vec{e}^{\rm T}J^{-2}\vec{e}$ 
are not easy to calculate in practical time. 
Hereafter, we call $\ve$ and $q_w$ of the optimal portfolio given in 
\sref{eq3} the minimal investment risk (per asset) and 
the concentrated investment level of the optimal portfolio, respectively, 
and we focus our analyze 
on these two characteristic quantities 
in order to 
evaluate 
the potential of an investment system.

With respect to 
the problem of computation size of determining the optimal portfolio, 
for the case that the return rates of assets are independently and identically distributed with mean and variance $0$ and $1$, 
respectively, 
$\ve$ and $q_w$ 
have already been analyzed theoretically and numerically 
using replica analysis 
in previous works\cite{Shinzato-Yasuda2015-BP,Shinzato2015-self-averaging}, in which the follow analytic formulations are reported: 
\bea
\ve\eq\f{\a-1}{2}\label{eq10},\\
q_w\eq\f{\a}{\a-1}\label{eq11},
\eea
where we use the ratio of the number of scenarios $p$ to 
the number of assets $N$, that is, 
the scenario rate $\a=p/N$. 
These formulations handle the $\a>1$ case ($p>N$) since in this case the optimal portfolio is uniquely determined.

In general, 
in an investment system, 
the return rates of assets in a securities market 
are not always independently and identically distributed, but rather can be correlated with each other and have different 
variances. Therefore, 
we need to implement asset management that appropriately 
combines assets in the securities market, including risk-free assets such as national government bonds and 
risky assets such as corporate bonds of small and medium-sized 
firms, unlike the approaches reported in previous works, which are not sufficient to handle 
the portfolio optimization problem 
when the variances of the return rates of assets are not identical.

Thus, our aim in this work is to analyze the optimal portfolio of 
the portfolio optimization problem with non-identical variances of the return 
rates of assets. For instance, 
investment assets are constructed to comprise risk-free assets and comparatively 
risky assets, since one can comparatively well characterize 
the nature of the actual market in this situation.
An additional aim is 
to develop a replica approach to solve this problem and to 
examine 
the typical behaviors of the minimal investment risk and the concentrated 
investment level of the optimal solution  in a 
way similar to that used in previous works. Specifically, in this paper, 
we assume that the return rates of assets are 
uncorrelated with one another and write 
the variance of asset $i$, 
 $E_X[x_{i\mu}^2]-(E_X[x_{i\mu}])^2=E_X[x_{i\mu}^2]$, 
as 
\bea
E_X[x_{i\mu}^2]
\eq s_i\label{eq12},
\eea
for simplicity. In addition, we assume that the $l$th moment of the return rate, 
$E_X[x_{i\mu}^l]$, does not depend on the number of assets and/or is 
finite, since it is known that 
the return rates of assets are not directly influenced by 
the asset size of the securities market.
\subsection{Boltzmann distribution}
In this subsection, we reformulate
the previously stated portfolio optimization problem using 
the Boltzmann distribution in the framework of 
probabilistic inference, similar to as was used in previous works,
in order to apply seamlessly the replica analysis approach.
Using the literature of probabilistic inference, 
the posteriori probability of portfolio $\vec{w}$ is defined as follows:
\bea
\label{eq13}P(\vec{w}|X)\eq\f{P_0(\vec{w})e^{-\b {\cal H}(\vec{w}|X)}}{Z(X)}
,
\eea
where the denominator in \sref{eq13} is defined as 
\bea
Z(X)\eq\area d\vec{w}P_0(\vec{w})e^{-\b{\cal H}(\vec{w}|X)},
\eea
which is called the partition function and plays a central role in 
the analysis of this paper. Moreover, 
the prior probability $P_0(\vec{w})$
is related in the budget constraint in \sref{eq2} and 
is exponentially described as $P_0(\vec{w})\propto \f{1}{(2\pi)^{\f{N}{2}}}
e^{k\left(\sum_{k=1}^Nw_i-N\right)}$, where 
auxiliary variable $k$ is chosen 
to satisfy  \sref{eq2}. In addition, 
$e^{-\b{\cal H}(\vec{w})}$ is a likelihood function or Boltzmann factor 
with inverse temperature $\b(>0)$, which is 
the control parameter in deriving the optimal solution. 

In practice, 
from the definition of the posteriori probability given in \sref{eq13}, 
because Boltzmann factor $e^{-\b{\cal H}}$ is a 
monotonic nonincreasing function with respect to 
${\cal H}$, the maximum a posteriori 
estimate is consistent with 
the optimal portfolio 
of the portfolio optimization problem.
In addition,
posteriori probability $P(\vec{w}|X)$ in 
a neighborhood of the maximum a posteriori 
estimate (the optimal portfolio of the portfolio optimization problem) increases as
inverse temperature $\b$ becomes large, whereas 
posteriori probability $P(\vec{w}|X)$ outside the neighborhood 
decreases to $0$ as $\b$ increases. 
In the limit of large $\b$, the expectation of ${\cal H}(\vec{w}|X)$, 
$E_{\vec{w}}[{\cal H}(\vec{w}|X)]=\area d\vec{w}P(\vec{w}|X){\cal H}(\vec{w}|X)$,
approaches 
\bea
\label{eq15}
\lim_{\b\to\infty}E_{\vec{w}}[{\cal H}(\vec{w}|X)]\eq
{\cal H}(\vec{w}^*|X)
\eea 
where $\vec{w}^*$ is the maximum a posteriori estimate. Moreover, this is related to the Boltzmann distribution as follows:
\bea
\label{eq16}
E_{\vec{w}}[{\cal H}(\vec{w}|X)]\eq
\area d\vec{w}P(\vec{w}){\cal H}
(\vec{w}|X)\nn
\eq-\pp{}{\b}\log Z(X),
\eea
which is known as the identical equation.
That is, 
from the properties in  \sref{eq15} and \sref{eq16}, 
the minimal investment risk per asset  $\ve$ is estimated using the following 
identical equation:
\bea
\ve\eq\f{1}{N}{\cal H}(\vec{w}^*|X)\nn
\eq\lim_{\b\to\infty}\left(-\pp{}{\b}\f{1}{N}\log Z(X)\right)
\label{eq17}.
\eea

\subsection{Replica analysis}
In order to examine the potential investment risk in our handled investment system,
we need to evaluate 
the expectation of minimal investment risk; however,  
it is well known that it is difficult to 
directly assess the expectation of $\ve$ defined in \sref{eq8} and/or 
the expectation of $\ve$ defined in \sref{eq17}.
Therefore, 
we 
will resolve this difficulty by using replica analysis in a way similar to that in
previous works. Specifically, since 
the expectation of $\ve$ defined in \sref{eq8}
is consistent with the expectation of $\ve$  defined in 
\sref{eq17}, 
we derive  $\ve=\lim_{\b\to\infty}\f{1}{N}E_X\left[E_{\vec{w}}\left[{\cal 
H}(\vec{w}|X)\right]\right]$ from $\f{1}{N}E_X\left[\log Z(X)\right]$,
although, 
unlike the specific model considered here, in general, 
directly averaging the logarithm of a random variable and/or statistic 
$Z(X)$ is not easy.
Therefore, we make use of either a replica trick, 
\bea
\label{eq18}\log Z(X)\eq\lim_{n\to0}\f{Z^n(X)-1}{n},
\eea
or the expectation description of the replica trick\cite{Nishimori}, 
\bea
\label{eq19}
E_X[\log Z(X)]
\eq
\lim_{n\to0}
\left\{
\begin{array}{l}
\f{E_X[Z^n(X)]-1}{n}\\
\f{\log E_X[Z^n(X)]}{n}\\
\pp{\log E_X[Z^n(X)]}{n}
\end{array}\right..
\eea
in order to 
analyze the expected value of the logarithm of the partition function. 
Thus, it is necessary to analytically assess $E_X[Z^n(X)]$ in the replica trick. 
We will need the following functions in order to 
perform this assessment using replica analysis:
\bea
\psi(n)\eq\lim_{N\to\infty}\f{1}{N}
\log E_X\left[Z^n(X)\right]\label{eq20},\\
\phi\eq\lim_{N\to\infty}\f{1}{N}E_X\left[\log Z(X)\right].
\eea
From these definitions, obviously, 
\bea
\label{eq21}
\phi\eq\lim_{n\to0}\pp{\psi(n)}{n},
\eea
holds and the minimal investment risk per asset $\ve$ is calculated as follows:
\bea
\label{eq22}
\ve\eq\lim_{\b\to\infty}\left(-\pp{\phi}{\b}\right),
\eea
where 
hereafter 
the expectation of 
the minimal investment risk per asset is 
also written simply as $\ve$ for convenience.

In replica analysis, our main {goal} is  to analyze 
$\psi(n)$ defined in \sref{eq20} in order to examine the typical behaviors 
of these statistics 
which characterize the investment 
systems. 
Determining these behaviors is easier 
than directly evaluating the expectation of the logarithm of 
the partition function; however, 
it is well known that it is hard to directly evaluate $\psi(n)$ ($n\in{\bf R}$), except for at specific values of $n$. 
On the other hand, since we first are possible to implement  
the configurational average of $Z^n(X)$ over the return rate matrix $X$
with respect to some $n\in{\bf N}$ 
using polynomial expansion of the partition function, 
we consider $E_X\left[Z^n(X)\right]$ only for $n\in{\bf N}$. It is hoped that the results for 
$n\in{\bf N}$ will provide cues for determining the  moment of the partition function at $n\in{\bf R}$.
Thus, for $n\in{\bf N}$, we will use the expansion
\bea
E_X\left[Z^n(X)\right]
\eq\area \prod_{a=1}^nd\vec{w}_aP_0(\vec{w}_a)\nn
&& E_X\left[
\exp
\left(
{-\b\sum_{a=1}^n{\cal H}(\vec{w}_a|X)}
\right)
\right],
\label{eq23}
\eea
where 
$\vec{w}_a=(w_{1a},w_{2a},\cdots,w_{Na})^{\rm T}\in{\bf 
R}^N,a=1,2,\cdots,n,$ is as defined previously above and $a$ is the replica index.
Moreover, since 
the return rate is independently drawn from a distribution with 
mean and variance $E_X[x_{i\mu}]=0$ and 
$E_X[x_{i\mu}^2]-(E_X[x_{i\mu}])^2=s_i$, respectively, and $\lim_{N\to\infty}N^{-\f{l}{2}}E_X[(x_{i\mu})^l]=0,(l=1,2,3,\cdots)$,
it is easy to implement the configurational averaging with 
the return distribution in our replica analysis 
in a way similar to that used in previous works\cite{Ciliberti,Shinzato2015-self-averaging}.

In order to calculate the right-hand side of \sref{eq23}, 
let us here define two types of 
{order parameters as follows:}
\bea
q_{wab}\eq\f{1}{N}\sum_{i=1}^Nw_{ia}w_{ib},\label{eq24}\\
q_{sab}\eq\f{1}{N}\sum_{i=1}^Nw_{ia}w_{ib}s_i\label{eq25},
\eea
where 
$a,b=1,2,\cdots,n$ are replica indices and $\tilde{q}_{wab}$ and 
$\tilde{q}_{sab}$ are {defined} as conjugate auxiliary variables of $q_{wab}$ and 
$q_{sab}$. Also, we define sets
$Q_w=\left\{q_{wab}\right\},\tilde{Q}_w=\left\{\tilde{q}_{wab}\right\},
Q_s=\left\{q_{sab}\right\},\tilde{Q}_s=\left\{\tilde{q}_{sab}\right\}\in
{\cal M}_{n\times n}
$. 
Note that 
for the case $a=b$,
\sref{eq24} corresponds to the concentrated investment level defined in 
\sref{eq4} and that 
we derive the order parameter {$q_{sab}$} and its conjugate auxiliary parameter 
$\tilde{q}_{sab}$ as the novel variables in order to handle 
the non-identical case (only $q_{wab},\tilde{q}_{wab}$ , and $k_a$ are used in 
our previous work\cite{Shinzato2015-self-averaging}).

Using the above variable definitions, 
\sref{eq23} can be expanded as follows:
\bea
\label{eq26}
\psi(n)
\eq
\mathop{\rm Extr}_{\Theta}
\left\{
-\vec{e}^{\rm T}\vec{k}
+\f{1}{2}\vec{k}^{\rm T}
\left\langle
\left(\tilde{Q}_w+s\tilde{Q}_s\right)^{-1}
\right\rangle\vec{k}
\right.
\nn
&&+\f{1}{2}{\rm Tr}Q_s\tilde{Q}_s
+\f{1}{2}{\rm Tr}Q_w\tilde{Q}_w
-\f{\a}{2}\log\det\left|
I+\b Q_s
\right|\nn
&&
\left.
-\f{1}{2}
\left\langle
\log \det\left|\tilde{Q}_w+s\tilde{Q}_s\right|
\right\rangle
\right\},
\eea
where $\vec{e}=(1,\cdots,1)^{\rm T}\in{\bf 
R}^n$ {is a vector of constants}, $\vec{k}=(k_1,\cdots,k_n)^{\rm 
T}\in{\bf R}^n$ {is an auxiliary vector}, $I\in{\cal M}_{n\times n}$ is the identity matrix, and 
$\Theta=\left\{\vec{k},Q_w,Q_s,\tilde{Q}_w,\tilde{Q}_s\right\}$, in which $Q_w, Q_s, \tilde{Q}_w$, and $\tilde{Q}_s$ are 
as defined above. Furthermore, the notation 
\bea
\left\langle f(s)
\right\rangle
\eq\lim_{N\to\infty}
\f{1}{N}
\sum_{i=1}^N
f(s_i)
\eea
is used and $\mathop{\rm Extr}_\Theta f(\Theta)$ means 
the extremum of the function $f(\Theta)$ with respect to $\Theta$. Moreover,
$p$ and $N$ 
are taken to approach infinity in order to maintain a finite scenario ratio 
$\a=p/N\sim O(1)$. Therefore, the extrema of the right-hand side of \sref{eq26} 
with respect to the order parameters are as follows: 
\bea
\pp{\psi(n)}{\vec{k}}
\eq-\vec{e}+
\left\langle
\left(\tilde{Q}_w+s\tilde{Q}_s\right)^{-1}
\right\rangle
\vec{k}=0,\\
\pp{\psi(n)}{Q_w}\eq\f{1}{2}\tilde{Q}_w=0,\\
\pp{\psi(n)}{Q_s}\eq\f{1}{2}\tilde{Q}_s-\f{\a\b}{2}
(I+\b Q_s)^{-1}
=0,\\
\pp{\psi(n)}{\tilde{Q}_w}
\eq-\f{1}{2}
\left\langle
\left(\tilde{Q}_w+s\tilde{Q}_s\right)^{-1}
\vec{k}
\vec{k}^{\rm T}
\left(\tilde{Q}_w+s\tilde{Q}_s\right)^{-1}
\right\rangle\nn
&&
-\f{1}{2}
\left\langle
\left(\tilde{Q}_w+s\tilde{Q}_s\right)^{-1}
\right\rangle+\f{1}{2}Q_w=0,\\
\pp{\psi(n)}{\tilde{Q}_s}
\eq-\f{1}{2}
\left\langle
s\left(\tilde{Q}_w+s\tilde{Q}_s\right)^{-1}
\vec{k}
\vec{k}^{\rm T}
\left(\tilde{Q}_w+s\tilde{Q}_s\right)^{-1}
\right\rangle\nn
&&
-\f{1}{2}
\left\langle
s\left(\tilde{Q}_w+s\tilde{Q}_s\right)^{-1}
\right\rangle+\f{1}{2}Q_s=0.
\eea
From these, 
the extrema of $Q_w$, $Q_s$, $\tilde{Q}_w$, $\tilde{Q}_s$, and $\vec{k}$ 
are determined as follows:
\bea
\label{eq33}
Q_w\eq
\f{\left\langle s^{-1}
\right\rangle}{\b(\a-1)}I+
\left(
\f{\left\langle s^{-2}
\right\rangle}{\left\langle s^{-1}
\right\rangle^2}
+
\f{1}{\a-1}
\right)D,
\\
Q_s\eq\f{1}{\b(\a-1)}I+\f{\a}{\left\langle s^{-1}
\right\rangle(\a-1)
}D,\\
\tilde{Q}_w\eq0,\\
\tilde{Q}_s\eq\b(\a-1)I-\f{\b^2(\a-1)}{\left\langle s^{-1}\right\rangle+n\b}D,\\
\vec{k}\eq\f{\b(\a-1)}{
\left\langle s^{-1}
\right\rangle
+n\b
}\vec{e},
\eea
where $D=\vec{e}\vec{e}^{\rm 
T}\in{\cal M}_{n\times n}$ is a square matrix with all components 1. Note that 
we do not require the ansatz of the replica symmetry solution, 
just as in our previous work\cite{Shinzato2015-self-averaging}.

Substituting these results into the right-hand side of \sref{eq26}, we analytically obtain
\bea
\label{eq39}
\psi(n)
\eq\f{n}{2}-\f{n}{2}
\left\langle
\log s
\right\rangle-\f{n\a}{2}\log\a+\f{n(\a-1)}{2}\log(\a-1)\nn
&&-\f{n}{2}\log\b-\f{\a-1}{2}\log\left(1+n\f{\b}{\left\langle s^{-1}\right\rangle}\right),
\eea
where 
 $\psi(0)=0$  is satisfied because $Z^0(X)=1$. 
Furthermore, 
although our analysis required the assumption $n\in{\bf N}$, 
the description of $\psi(n)$ given in \sref{eq39} 
can be applied more generally. Namely,  
we assume that 
\sref{eq39} has an analytic continuation to $n\in{\bf R}$ 
and substitute it into  \sref{eq21}, giving
\bea
\phi\eq\f{1}{2}-\f{1}{2}\left\langle
\log s
\right\rangle
-\f{\a}{2}\log\a+\f{\a-1}{2}\log(\a-1)\nn
&&-\f{1}{2}\log\b-\f{\b(\a-1)}{2\left\langle s^{-1}
\right\rangle
}.
\eea
Furthermore, the minimal investment risk per asset $\ve$ is 
obtained from \sref{eq22} as follows:
\bea
\ve\eq\f{\a-1}
{2\left\langle s^{-1}
\right\rangle
}\label{eq41}.
\eea
In a similar way, since concentrated investment level $q_w$ is evaluated 
 by using $a=b$ in \sref{eq24}, from the diagonal of $Q_w$ 
 in \sref{eq33},
\bea
q_w\eq
\f{\left\langle s^{-2}
\right\rangle
}
{\left\langle s^{-1}
\right\rangle^2
}+\f{1}{\a-1}\label{eq42},
\eea
can be analytically evaluated  where 
the concentrated investment level of the optimal portfolio is 
derived from $q_w$ in the limit of sufficiently large $\b$.

One last point should be noted here.
It remains an open problem 
whether 
the result for 
the $n\in{\bf N}$ case 
{extends to $n\in{\bf R}$ as an analytic continuation.} Furthermore, 
we need to verify the effectiveness of 
our assumption, so we will compare the findings of our proposed method with those of numerical experiments below. 
\subsection{Interpretation of our findings of two feature quantities}
Before the verification of 
the effectiveness of our assumption and our proposed approach by using numerical experiments, 
we give the interpretation of 
the results of our replica analysis. 
\paragraph{(a) Two feature quantities depend only on the variance of the return rate 
    distribution }
From 
\sref{eq41} and 
\sref{eq42}, 
it is clear that 
these statistics do not depend on the other details of 
the return rate distribution besides the variances of the return rates of the assets. 
Suppose one rescales the return rate, such as in a stock split, 
with respect to a scaling coefficient of return $\sqrt{\g}$, that is,
the rescaled return rate $\bar{x}_{i\mu}=\sqrt{\g}x_{i\mu}$ is defined from the original return rate $x_{i\mu}$.
Then $E_X[\bar{x}_{i\mu}]=0
$ and $E_X[\bar{x}^2_{i\mu}]=\g s_i$ are satisfied 
in the rescaled investment system, and the minimal investment risk is described by 
the following: 
\bea
\ve(\g)\eq
\lim_{N\to\infty}
\f{1}{2N}
\sum_{\mu=1}^p
\left(\f{1}{\sqrt{N}}
\sum_{i=1}^N\bar{x}_{i\mu}w_i
\right)^2,
\eea
and the concentrated investment level of the optimal portfolio is
\bea
q_w(\g)\eq\lim_{N\to\infty}
\f{1}{N}\sum_{i=1}^Nw_i^2.
\eea
From replica analysis, 
these statistics  are estimated as follows:
\bea
\ve(\g)\eq\f{\g(\a-1)}{2\left\langle s^{-1}
\right\rangle
},\\
q_w(\g)\eq
\f{\left\langle s^{-2}
\right\rangle
}
{\left\langle s^{-1}
\right\rangle^2
}+\f{1}{\a-1}.
\eea
Since the size of 
the portfolio of each asset is not changed by scaling coefficient $\sqrt{\g}$ 
under budget constraint \sref{eq2}, 
the concentrated investment level is considered to be unchanged  and 
the minimal investment risk changes by a factor of $\sqrt{\g}$.

\paragraph{(b) Comparison with the findings in previous works}
If the variances of the return rates of assets are identical, 
for instance, $s_i=1$, then 
$\left\langle s^{-1}
\right\rangle=\left\langle s^{-2}
\right\rangle=1
$, and thus our results agree with \sref{eq10} and \sref{eq11} reported in previous works\cite{Shinzato-Yasuda2015-BP,Shinzato2015-self-averaging}.
\paragraph{(c) Lower bound on concentrated investment level}
For the lower bound on the concentrated investment level, 
from the relation 
$\left\langle s^{-2}\right\rangle-\left\langle s^{-1}\right\rangle^2
=\left\langle
\left( s^{-1}-\left\langle s^{-1}\right\rangle\right)^2
\right\rangle
\ge0
$, the relation
$q_w\ge1+\f{1}{\a-1}=\f{\a}{\a-1}$ can be obtained. 
Moreover, 
regarding the sharp lower bound, 
because $\left\langle s^{-2}\right\rangle-\left\langle 
s^{-1}\right\rangle^2=0$, the variances of the return rates of the assets are identical.

\paragraph{(d) Comparison with operations research findings}
In the present paper, 
we have found the optimal portfolio minimizing an investment risk ${\cal H}(\vec{w}|X)$ defined by
a given return rate matrix, $X=\left\{\f{x_{i\mu}}{\sqrt{N}}\right\}\in{\cal 
M}_{N\times p}$, and have determined 
the typical behaviors of two features of an optimal portfolio; in other 
words, 
we analyzed the quenched disorder  system of this portfolio optimization problem 
discussed in the literature of 
statistical mechanical informatics\cite{Shinzato2015-self-averaging}. For comparison, 
we will also analyze this optimal problem using the standard approach in operations research.
In this approach, one first averages the investment risk ${\cal H}(\vec{w}|X)$ with the whole configuration 
of return rate matrix $X$. 
Next, one finds the optimal portfolio minimizing 
the expected investment risk $E_X\left[{\cal H}(\vec{w}|X)\right]$. 
The procedure of this approach is consistent with an annealed disorder  system approach in the context of statistical mechanical informatics\cite{Shinzato2015-self-averaging,Nishimori}.

Now, let us evaluate $\ve$ and 
$q_w$ of the optimal portfolio in the annealed disorder  system
 (the standard approach in operations research). From the statistical properties of the defined return rate, since 
$E_X\left[x_{i\mu}^2\right]=s_i$ and 
$E_X\left[x_{i\mu}x_{j\mu}\right]=E_X[x_{i\mu}]E_X[x_{j\mu}]=0,(i\ne j)$
are assumed above, 
the expected investment risk in the annealed disorder  system is obtained as follows:
\bea
E_X\left[{\cal H}(\vec{w}|X)\right]
\eq\f{\a}{2}
\sum_{i=1}^Nw_i^2s_i\label{eq47},
\eea
Furthermore, we find the optimal portfolio 
under the budget constraint in \sref{eq2}, {from which}
\bea
w_i^{\rm OR}\eq\f{s_i^{-1}}{\left\langle s^{-1}\right\rangle},
\eea
is obtained. This result implies that 
the portfolio is proportional to the inverse of the variance of the return rate of the asset.
Finally, the concentrated investment level of the optimal portfolio minimizing the expected 
investment risk is briefly calculated as follows:
\bea
q_w^{\rm OR}\eq\f{\left\langle s^{-2}\right\rangle}
{\left\langle s^{-1}\right\rangle^2}.
\eea
This implies that the first term of the concentrated investment 
level in the quenched disorder  system on the right-hand side of \sref{eq42}
is caused by the annealed disorder  system. As an interpretation of this relationship, 
since the return rates of assets are assumed to be independently distributed, 
there is no correlation between asset $i$ and asset $j$ in the expectation of the investment risk in 
\sref{eq47}. 
In the other words, 
this investment system is 
equivalent to a securities market comprising $N$ assets which are 
uncorrelated with each other. 
However, in general, 
when 
deriving the solution which minimizes
an investment risk defined by a given random return rate 
matrix 
$X$ in a quenched disorder  system, the correlation terms ignored in 
the annealed disorder  system cannot be ignored and so the second term on the right-hand side of \sref{eq42}, $\f{1}{\a-1}$, remains.
Moreover, from another viewpoint, 
when $\a\simeq1$,
the optimal solution is close to  
the eigenvector of the minimal eigenvalue of $J=XX^{\rm T}\in{\cal 
M}_{N\times N}$, 
$\l_{\min}$. Since $\l_{\min}$ is close to $0$ for $\a\simeq1$ and 
$q_w\simeq\f{1}{\l_{\min}^2}$, 
the concentrated investment level $q_w$ becomes infinity. 
Since the number of reference scenarios is {relatively} small, 
one interpretation is that an investor should concentrate investments in 
relatively riskless assets, based on the return rate table (like the previously mentioned concentrated investment strategy). 
 On the other hand, when 
$\a\gg1$, since the number of reference scenarios is 
{large enough}, 
the correlation term $x_{i\mu}x_{j\mu}$ is close to $0$ 
relative to the autocorrelation term $x_{i\mu}^2$, 
the investment risk can be considered to be 
well approximated by the
expected investment risk in the annealed disorder  system, $E_X\left[{\cal 
H}(\vec{w}|X)\right]$.
Furthermore, the behavior of 
the concentrated investment level is similar to that in an annealed disorder  system.

Similarly, the minimal expected investment risk in the annealed disorder  system is 
briefly calculated as follows:
\bea
\ve^{\rm OR}\eq\f{\a}{2\left\langle s^{-1}\right\rangle}
\label{eq50}.
\eea
If we compare the results of the quenched disorder  system 
in 
\sref{eq41} with those of the annealed disorder system in \sref{eq50}, 
it is clear that the effect of correlation between assets {is captured 
by} $-\f{1}{2\left\langle s^{-1}\right\rangle}$.

\section{Discussion}
In this section, we 
discuss 
the effectiveness  
of 
our proposed method by comparing its results with those from numerical experiments.
\subsection{Case 1: Two types of variances}
First, 
we consider here 
the portfolio optimization problem with 
two possible variances of the return rate; this problem relates to the case of 
risk-free and risky assets. Namely, 
we assign the variance of the return rate of asset $i$, $s_i$, as $1$ with probability $r$ 
and 
as $\tilde{s}$ with probability $1-r$.
Then, the expectation of $s^{-t}$, $\left\langle s^{-t}\right\rangle$, is 
calculated as follows:
\bea
\left\langle s^{-t}\right\rangle\eq r+(1-r)\tilde{s}^{-t},\qquad(t=1,2).
\eea
We implement numerical experiments using the following three numerical 
settings: 
(A) $\left\langle s^{-1}\right\rangle=3,
\left\langle s^{-2}\right\rangle=30$, that is, 
	   $r=\f{21}{25},\tilde{s}=\f{2}{27}$.
(B) $\left\langle s^{-1}\right\rangle=4,
\left\langle s^{-2}\right\rangle=30$, that is, 
	   $r=\f{14}{23},\tilde{s}=\f{3}{26}$.
(C) $\left\langle s^{-1}\right\rangle=5,
\left\langle s^{-2}\right\rangle=30$, that is, 
	   $r=\f{5}{21},\tilde{s}=\f{4}{25}$. 
Here, we solve for the portfolio which minimizes the investment risk ${\cal H}(\vec{w}|X)$
defined by 
a random return rate matrix $X$ whose 
entries are independently distributed with mean and 
variance $0$ and $s_i$, respectively, 
by using the optimization algorithm based on the steepest descent method shown in Fig \ref{fig1}. 
This algorithm 
finds the extremum of Lagrange function $L(\vec{w},\zeta)={\cal 
H}(\vec{w}|X)+\zeta(N-\vec{e}^{\rm T}\vec{w})$ with respect to portfolio 
$\vec{w}$ and auxiliary variable $\zeta$, 
where $\vec{e}=\left\{1,1,\cdots,1\right\}^{\rm T}\in{\bf R}^N$.

The results of the algorithm based on the steepest descent method 
(symbols with error bars),  
 the replica analysis (solid lines), and 
the standard approach in operations research (dashed lines) are shown 
in Figs. \ref{fig2} and \ref{fig3}. 
In both figures, results appear in the 
order (A), (B), and (C) from top to bottom. 
More specifically, 
the symbols with error bars
show 
the expectation of the minimal investment risk 
derived by numerical simulations, in which the number of assets is set as $N=1000$, 
and we randomly generate $100$ return rate matrices and solve 
for the portfolio which minimizes the investment risk ${\cal 
H}(\vec{w}|X)$ by using the steepest descent 
 method introduced in 
Fig. \ref{fig1}. 

Fortunately, these figures show that the typical behaviors of 
minimal investment risk derived by numerical experiment 
are in agreement with those derived by replica analysis. 
On the other hand, these figures also show that the standard approach in operations 
research is too difficult to analyze the 
potential of the optimization problem. 
Namely, it is found that our proposed  method based on statistical mechanical informatics 
 can assess the typical behaviors of the optimal solution minimizing this 
 portfolio optimization problem 
without the difficulty of the standard approach.

\begin{figure}[tbh]
\begin{description}
\item[Step 0] At $t=0$, 
set the initial state as
$\vec{w}_0=\vec{e}$ and $\zeta_0=1$. Two small positive step sizes 
$\eta_w,\eta_\zeta>0$ are determined and 
a small positive constant number 
$\d>0$ which is required in the stopping condition is set. In this study, $\eta_w=100/N,\eta_\zeta=1/N,\d=10^{-6}$ are used.
\item[Step 1]
Using $\vec{w}_{t},\zeta_{t}$, we 
determine $\vec{w}_{t+1},\zeta_{t+1}$ as follows: 
\bea
\vec{w}_{t+1}\eq\vec{w}_t-\eta_w\left(XX^{\rm T}\vec{w}_t-\zeta_t\vec{e}\right),\\
\zeta_{t+1}\eq\zeta_t+\eta_\zeta\left(N-\vec{e}^{\rm T}\vec{w}_t\right).
\eea
\item[Step 2]
If the L$1$ distance between 
$\vec{w}_{t+1},\zeta_{t+1}$ and $\vec{w}_t,\zeta_t$, 
$\Delta=|\zeta_{t+1}-\zeta_t|+\sum_{k=1}^N\left|w_{k,t+1}-w_{k,t}\right|
$, is greater than $\d$, then 
$t\to t+1$ and go to 
{\bf Step 1}; 
otherwise, go to {\bf Step 3}.
\item[Step 3]
Return the results as the optimal portfolio (or its approximate value).
\end{description}
\caption{\label{fig1}
Iterative algorithm for optimizing based on the steepest descent method. 
}
\end{figure}

\begin{figure}[bht]
\begin{center}
\includegraphics[width=0.6\hsize,angle=270]{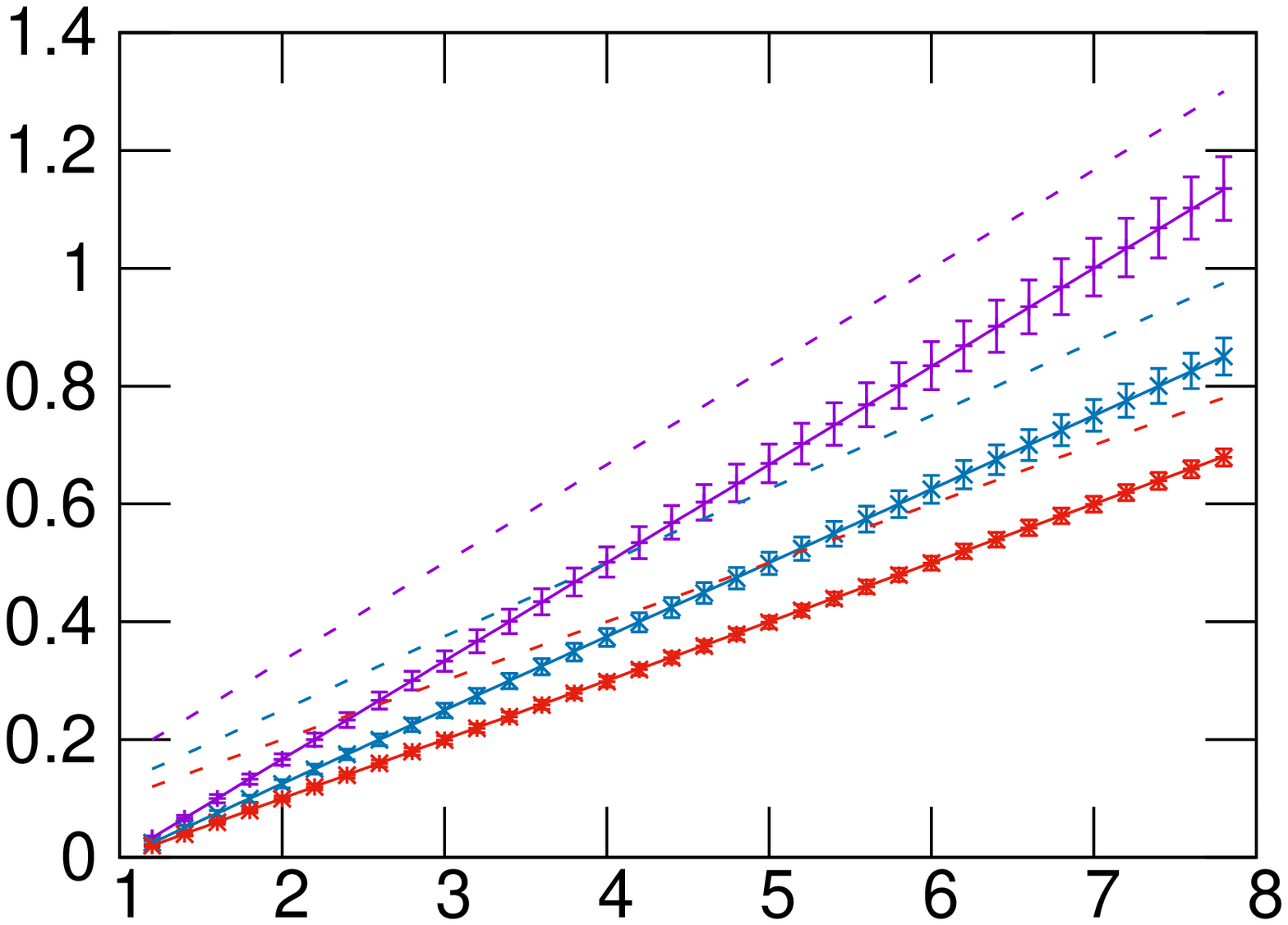} 
\caption{\label{fig2}
Minimal investment risk per asset $\ve$ versus scenario ratio $\a=p/N$ 
 from a numerical experiment (symbols with error bars), our proposed 
 approach (solid lines), and a standard operations research approach 
 (dashed lines) for Case 1. 
Colors correspond to Case 1(A) (purple), 
Case 1(B) (blue), and Case 1(C) (red). 
}
\end{center}
\begin{center}
\includegraphics[width=0.6\hsize,angle=270]{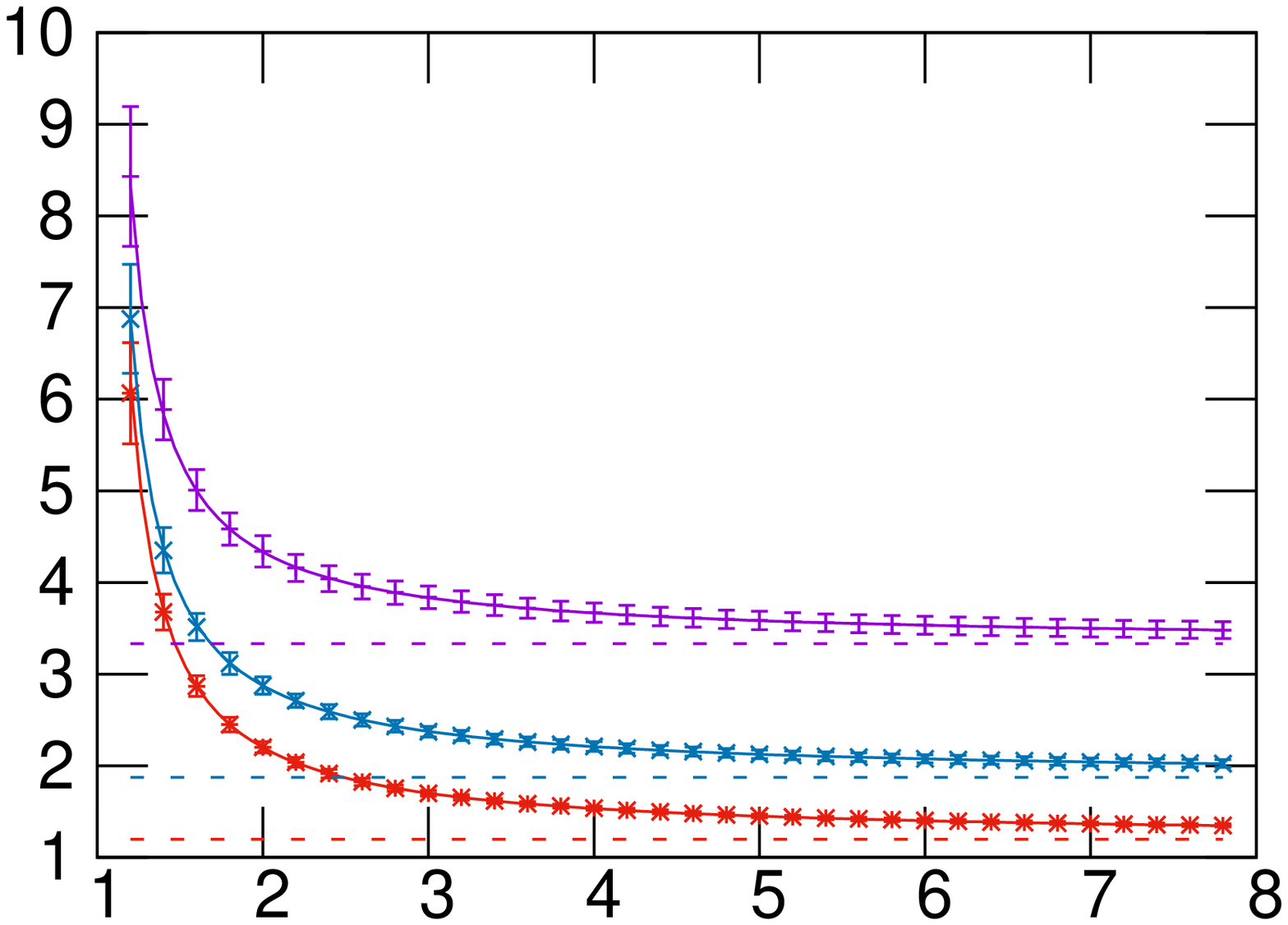} 
\caption{\label{fig3}
Concentrated investment level of optima portfolio $q_w$ versus scenario ratio $\a=p/N$ 
 from a numerical experiment (symbols with error bars), our proposed 
 approach (solid lines), and a standard operations research approach 
 (dashed lines) for Case 1. 
Colors correspond to Case 1(A) (purple), 
Case 1(B) (blue), and Case 1(C) (red).
}
\end{center}
\end{figure}

\subsection{Case 2: Variance is distributed uniformly}
\begin{figure}[b]
\begin{center}
\includegraphics[width=0.6\hsize,angle=270]{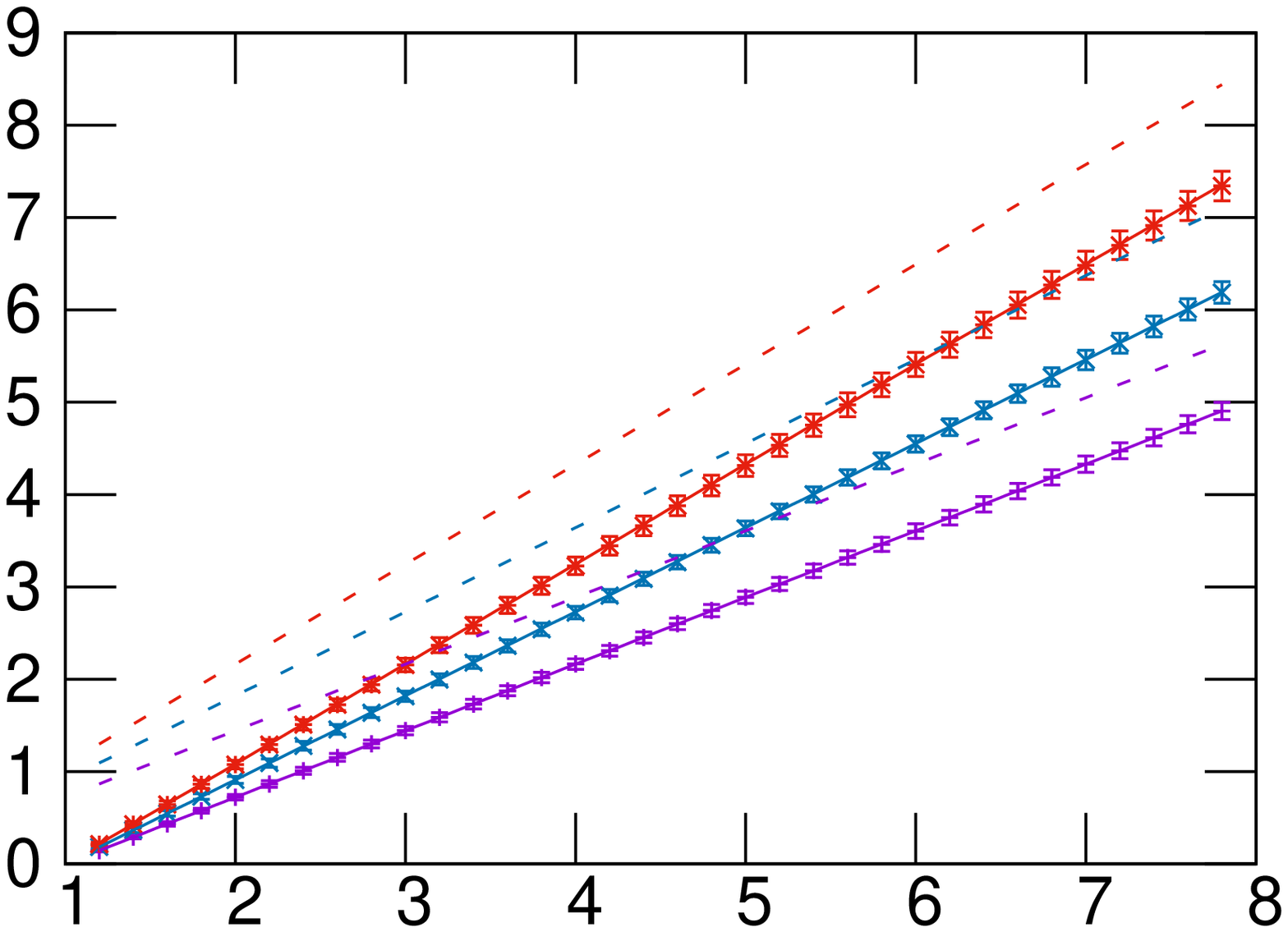} 
\caption{\label{fig4}
Minimal investment risk per asset $\ve$ versus scenario ratio $\a=p/N$ 
 from a numerical experiment (symbols with error bars), our proposed 
 approach (solid lines), and a standard operations research approach 
 (dashed lines) for Case 2. 
Colors correspond to Case 2(A') (purple), 
Case 2(B') (blue), and Case 2(C') (red). 
}
\end{center}
\begin{center}
\includegraphics[width=0.6\hsize,angle=270]{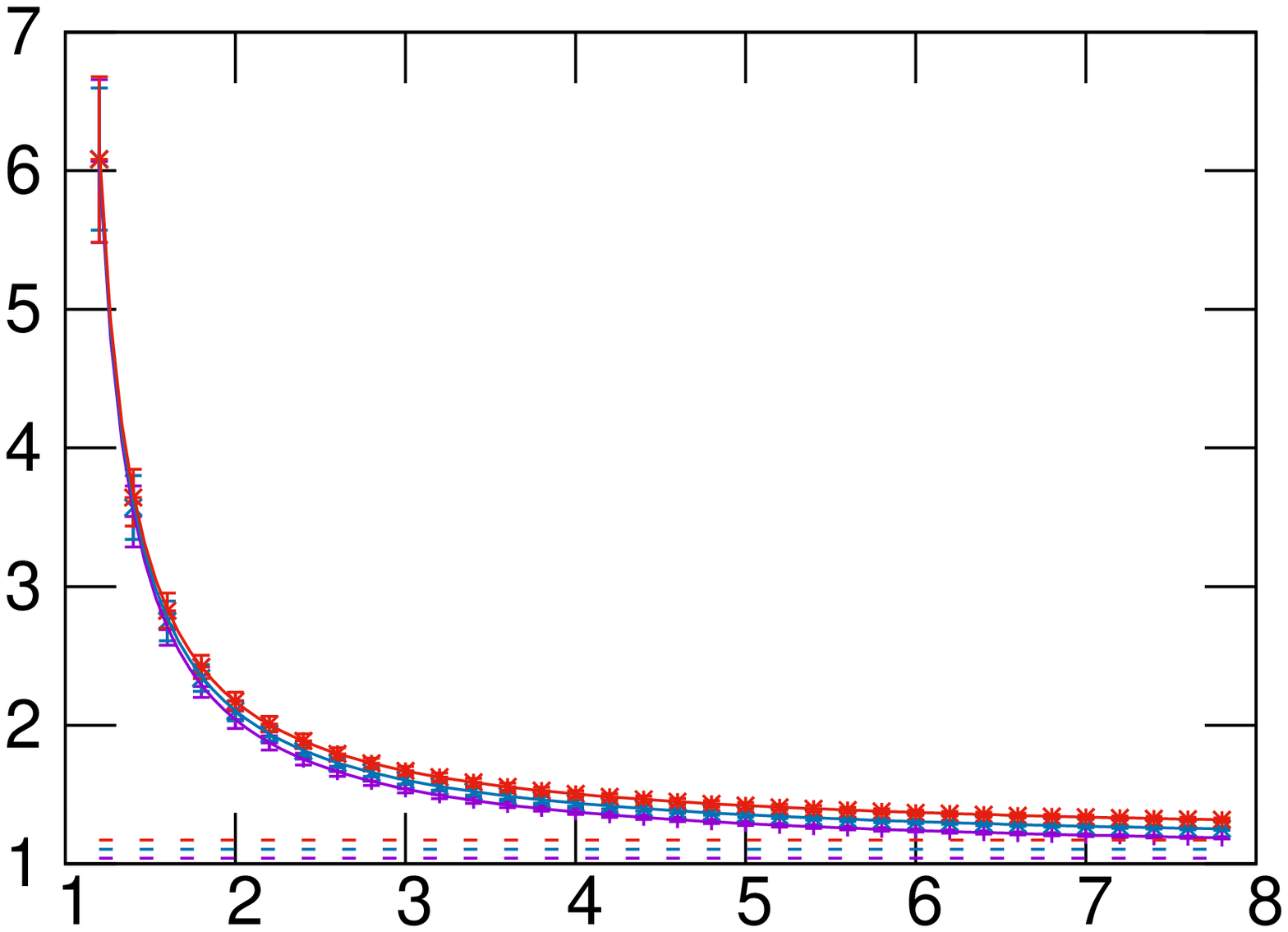} 
\caption{\label{fig5}
Concentrated investment level of optima portfolio $q_w$ versus scenario ratio $\a=p/N$ 
 from a numerical experiment (symbols with error bars), our proposed 
 approach (solid lines), and a standard operations research approach 
 (dashed lines) for Case 2. 
Colors correspond to Case 2(A') (purple), 
Case 2(B') (blue), and Case 2(C') (red). 
}
\end{center}
\end{figure}

Next, we discuss
the case that 
the variance of the return rate has 
a continuous uniform distribution.
Namely, 
the range of the variance of the return rate of asset $i$, $s_i$, 
is given by $l_s\le s_i\le u_s$.
Then 
$\left\langle
s^{-1}
\right\rangle
=\f{\log\f{u_s}{l_s}}{u_s-l_s}
$ and 
$\left\langle
s^{-2}
\right\rangle
=\f{1}{u_sl_s}
$ are obtained and 
the minimal investment risk per asset and concentrated investment level are 
as follows:
\bea
\ve\eq\f{\a-1}{2}
\f{u_s-l_s}{\log\f{u_s}{l_s}},
\\
q_w\eq
\f{1}{u_sl_s}\left(\f{u_s-l_s}{\log\f{u_s}{l_s}}\right)^2
+
\f{1}{\a-1}.
\eea
In this case, we implement numerical simulations 
with the following three settings: (A') $l_s=1,u_s=2$, 
(B') $l_s=1,u_s=3$, and (C') $l_s=1,u_s=4$. 
For the numerical simulations, the number of assets is set as $N=1000$, 
and we randomly generate $100$ return rate matrices and solve 
for the portfolio which minimizes the investment risk by using the steepest descent 
 method introduced in 
Fig. \ref{fig1}. The results are shown in Figs. \ref{fig4}
 and \ref{fig5}. 
In both figures, results appear in the 
order (A'), (B'), and (C') from bottom to top. 
Similar to in Case 1, 
the typical behaviors of 
the minimal investment risk derived by numerical experiments 
and the results derived by replica analysis 
 are in good agreement, fortunately. 
On the other hand, these figures again show that the standard approach in operations 
research is too difficult to analyze the 
potential of the optimization problem. 
Namely, it is found again that our proposed  method based on statistical mechanical informatics 
 can assess the typical behaviors of the optimal solution minimizing this 
 portfolio optimization problem 
without the difficulty of the standard approach.

\section{Conclusion}
In the present work, 
the portfolio optimization problem 
in which the variances of the return rates of assets are not identical,
for instance, risk-free assets represented by national government bonds 
and risky assets represented by the corporate bonds of small and 
medium-sized firms, has 
been analyzed. The analysis in particular focused on two features, 
minimal investment risk and 
concentrated investment level, which were analyzed using replica analysis. 
Moreover, 
we discussed the 
potential of replica analysis developed in  statistical mechanical informatics 
in resolving the portfolio optimization problem. 
The approach, 
which was 
developed by Shinzato et al. in previous works under the assumption of 
identical variances of the return rates of 
assets, is improved upon in the current work to allow it to
solve the portfolio optimization problem with non-identical 
variances. 
With the improved method, the behavior of the optimal portfolio in 
the portfolio optimization problem can be 
derived using replica analysis.
It was shown that if the scenario ratio is close to $1$, then 
the concentrated investment strategy gives the optimal portfolio, whereas if the scenario ratio is 
sufficiently large, then 
the equipartition investment strategy gives the optimal portfolio. In addition, 
we verified  
the effectiveness of our proposed method through 
numerical experiments.

In our future work, 
based on our assumption 
that the return rates of assets are independently distributed, for simplicity,
though we developed the replica approach in previous works, 
for the actual securities market,
we should consider correlation between assets, which the present work also ignored. Therefore, to analyze the nature of a real securities market,
the minimal investment risk and concentrated investment level 
in the portfolio optimization problem in which returns are correlated with 
each other should be analyzed.
Obviously, the proposed improvement in the present work 
does not allow 
analysis of a correlated investment system; for the correlated case, 
we need to develop an appropriate replica analysis approach.

\section{Acknowledgments}
The author thanks I. Kaku and T. Mizuno for their fruitful comments.
Also, the author is grateful to R. Wakai and K. Kobayashi for valuable discussions.
The work is partly supported by Grant-in-Aid Nos. 24710169 and 15K20999, 
President project for young scientists in Akita prefectural university, 
Research Project No. 50 in the National Institute of Informatics, Japan, 
Research Project No. 5 in Japan Institute of Life Insurance, 
Kyoto university institute of economic research foundation, 
Research Project No. 1414 in Zengin Foundation for Studies on Economics 
and Finance, 
Research Project No. 2068 in The institute of statistical mathematics, 
and Research Project No. 2 in Kampo foundation.


\section{Appendix}

\subsection{Replica analysis using replica symmetry ansatz}
In this appendix, 
using the replica symmetry ansatz, 
we analyze 
the behaviors of 
two features of the optimal portfolio 
in a way similar to that in previous works\cite{Ciliberti,Nishimori}.  In general, 
we cannot analyze 
$\psi(n)$ for an arbitrary $n\in{\bf R}$ directly, 
because we cannot differentiate the right-hand side of \sref{eq26} with respect to the 
replica number 
$n$. Thus, 
as mentioned in the above main manuscript, $n\in{\bf N}$ is employed for simplicity.
Moreover, it is necessary to assume the optimal solution descriptions  of 
parameters $q_{wab},q_{sab},\tilde{q}_{wab},\tilde{q}_{sab},k_a$ in 
\sref{eq26} appropriately. 
{Specifically, the parameters are assumed to be} as follows: 
\bea
\label{eq56}
q_{wab}\eq\left\{\begin{array}{ll}
\chi_{w}+q_w&a=b\\
q_w&a\ne b
\end{array}
\right.,\\
q_{sab}\eq\left\{\begin{array}{ll}
\chi_{s}+q_s&a=b\\
q_s&a\ne b
\end{array}
\right.,\\
\tilde{q}_{wab}\eq\left\{\begin{array}{ll}
\tilde{\chi}_{w}-\tilde{q}_w&a=b\\
-\tilde{q}_w&a\ne b
\end{array}
\right.,\\
\tilde{q}_{sab}\eq\left\{\begin{array}{ll}
\tilde{\chi}_{s}-\tilde{q}_s&a=b\\
-\tilde{q}_s&a\ne b
\end{array}
\right.,\\
k_a\eq k.
\eea
This solution is called the replica symmetry solution and the above assumption is 
called the replica symmetry ansatz. 
The effectiveness of the replica symmetry solution and that of the replica symmetry assumption have been verified 
in the analyses of several problems in statistical learning theory and 
information theory\cite{Shinzato2008,Kabashima2004}. 
In this paper, 
for the sake of simplicity, the replica symmetry solution is used,  where 
from {the expressions} of $q_w$ and $\chi_w$,
since there exists symmetry in the replica indices $a,b$,
$E_{\vec{w}}[w_{ia}^2]=E_{\vec{w}}[w_i^2]$ and 
$E_{\vec{w}}[w_{ia}w_{ib}]=E_{\vec{w}}[w_{ia}]E_{\vec{w}}[w_{ib}]
=(E_{\vec{w}}[w_i])^2,(a\ne b)$ are used. Therefore, 
\bea
q_{waa}\eq\lim_{N\to\infty}
\f{1}{N}
\sum_{i=1}^NE_{\vec{w}}[w_{i}^2],\\
q_{wab}
\eq\lim_{N\to\infty}
\f{1}{N}
\sum_{i=1}^N(E_{\vec{w}}[w_{i}])^2,
\eea
are obtained. Especially, $\chi_w=q_{waa}-q_{wab}=q_{waa}-q_w$ from \sref{eq56} is 
\bea
\chi_w\eq\lim_{N\to\infty}
\f{1}{N}
\sum_{i=1}^N
\left(
E_{\vec{w}}[w_i^2]
-(E_{\vec{w}}[w_i])^2
\right).
\eea
Now, from the replica symmetry ansatz, 
\bea
\psi(n)\eq-\f{n\a}{2}\log(1+\b\chi_s)-\f{\a}{2}\log\left(1+\f{n\b q_s}{1+\b\chi_s}\right)\nn
&&+\f{n}{2}(\chi_w+q_w)(\tilde{\chi}_w-\tilde{q}_w)-\f{n(n-1)}{2}q_w\tilde{q}_w\nn
&&+\f{n}{2}(\chi_s+q_s)(\tilde{\chi}_s-\tilde{q}_s)-\f{n(n-1)}{2}q_s\tilde{q}_s\nn
&&-nk +\f{1}{2}\left\langle
\f{nk ^2}{\tilde{\chi}_w+s\tilde{\chi}_s-n(\tilde{q}_w+s\tilde{q}_s)
}
\right.
\nn
&&
\left.
-n\log(\tilde{\chi}_w+s\tilde{\chi}_s)
-\log\left(1-n\f{\tilde{q}_w+s\tilde{q}_s}{\tilde{\chi}_w+s\tilde{\chi}_s}\right)
\right\rangle,\nn
\label{eq64}
\eea
is obtained. 
From \sref{eq21} and \sref{eq64},
\bea
\phi
\eq-\f{\a}{2}\log(1+\b\chi_s)-\f{\a\b q_s}{2\left(1+\b\chi_s\right)}+\f{1}{2}q_w\tilde{q}_w\nn
&&+\f{1}{2}(\chi_w+q_w)(\tilde{\chi}_w-\tilde{q}_w)
+\f{1}{2}(\chi_s+q_s)(\tilde{\chi}_s-\tilde{q}_s)\nn
&&+\f{1}{2}q_s\tilde{q}_s-k +\f{1}{2}\left\langle
\f{k ^2}{\tilde{\chi}_w+s\tilde{\chi}_s
}
-\log(\tilde{\chi}_w+s\tilde{\chi}_s)
\right.\nn
&&\left.
+\f{\tilde{q}_w+s\tilde{q}_s}{\tilde{\chi}_w+s\tilde{\chi}_s}
\right\rangle\label{eq65}
\eea
is obtained. Furthermore, 
we need to determine the extremum  
of order parameters 
$\chi_w,q_w,\chi_s,q_s$ and their conjugate 
order parameters $\tilde{\chi}_w,\tilde{q}_w,
\tilde{\chi}_s,\tilde{q}_s,k $
in 
the right-hand side of \sref{eq65}. From the 
extremum of the right-hand side of 
\sref{eq65},
\bea
\label{eq66}
\chi_w\eq\f{\left\langle s^{-1}\right\rangle}{\b(\a-1)},\\
\label{eq67}q_w\eq\f{\left\langle s^{-2}\right\rangle}
{\left\langle s^{-1}\right\rangle^2}+\f{1}{\a-1},\\
\tilde{\chi}_w\eq0,\\
\tilde{q}_w\eq0,\\
\chi_s\eq\f{1}{\b(\a-1)},
\eea

\bea
q_s\eq\f{\a}{\left\langle s^{-1}\right\rangle(\a-1)},\\
\tilde{\chi}_s\eq\b(\a-1),\\
\tilde{q}_s\eq\f{\b^2(\a-1)}{\left\langle s^{-1}\right\rangle},\\
k\eq\f{\b(\a-1)}{\left\langle s^{-1}\right\rangle},
\eea
are obtained. Then, 
the investment risk per asset of finite temperature $\b$ is 
described as follows:
\bea
\ve\eq\f{\a-1}{2}\left(\f{1}{\left\langle s^{-1}\right\rangle}+\f{1}{\b(\a-1)}\right).
\eea
Moreover, in the limit of large $\b$, 
\bea
\ve\eq\f{\a-1}{2\left\langle 
s^{-1}\right\rangle},
\eea
is obtained and the concentrated investment level is also assessed in \sref{eq67}.
\subsection{Belief propagation algorithm}
In our previous work\cite{Shinzato-Yasuda2015-BP}, 
an algorithm for solving the portfolio optimization problem defined in \sref{eq1} and \sref{eq2}
based on belief propagation was proposed. 
Note that the statistical nature of the return rate 
was not limited in our previous work, that is, 
we do not assume that the return rate of each asset is independently and 
identically distributed. 
 Thus, let 
the findings in this paper be supported by this belief propagation 
algorithm. First, this algorithm is expanded as follows:
\bea
\label{eq77}
m_{wi}\eq \chi_{wi}(h_{wi}+k),\\
h_{wi}\eq
\f{1}{\sqrt{N}}
\sum_{\mu=1}^px_{i\mu}m_{u\mu}+\tilde{\chi}_{wi}m_{wi},\\
\tilde{\chi}_{wi}\eq\f{1}{N}
\sum_{\mu=1}^px_{i\mu}^2\chi_{u\mu},\\
\chi_{wi}\eq\f{1}{\tilde{\chi}_{wi}},\\
m_{u\mu}\eq-\chi_{u\mu}h_{u\mu},\\
h_{u\mu}\eq\f{1}{\sqrt{N}}
\sum_{i=1}^Nx_{i\mu}m_{wi}-\tilde{\chi}_{u\mu}m_{u\mu},\\
\tilde{\chi}_{u\mu}\eq\f{1}{N}
\sum_{i=1}^Nx_{i\mu}^2\chi_{wi},\\
\label{eq84}\chi_{u\mu}\eq\f{\b}{1+\b\tilde{\chi}_{u\mu}},
\eea
where $m_{wi}=E_{\vec{w}}[w_i]$ and 
$\chi_{wi}=E_{\vec{w}}[w_i^2]-(E_{\vec{w}}[w_i])^2$
are already used in \cite{Shinzato-Yasuda2015-BP} {and it 
is indicated that the solution of these simultaneous equations 
is consistent with the optimal portfolio in the previous work. }
In practice, from 
\sref{eq77} to \sref{eq84},
\bea
k\eq\f{\b}{N}
\sum_{j=1}^N
\sum_{\mu=1}^px_{i\mu}x_{j\mu}m_{wj},
\eea
is obtained. This can be expressed in the following vector-matrix form:
\bea
k\vec{e}\eq\b J\vec{m}_w,\label{eq86}
\eea
where matrix $J=\left\{J_{ij}\right\}\in{\cal M}_{N\times N}$ is given in \sref{eq4} and $\vec{m}_w
=(m_{w1},m_{w2},\cdots,m_{wN}
)^{\rm T}\in{\bf R}^N
$. 
This implies $\vec{m}_w=\f{k}{\b}J^{-1}\vec{e}$.
In addition, since $\vec{m}_w$ satisfies \sref{eq2}, then 
$\f{k}{\b}=\f{N}{\vec{e}^{\rm T}J^{-1}\vec{e}}$ is obtained.
If we substitute this into \sref{eq86}, then 
\bea
\vec{m}_w\eq
\f{NJ^{-1}\vec{e}}{\vec{e}^{\rm T}J^{-1}\vec{e}},
\eea
{is obtained, which can easily be shown to be} consistent with \sref{eq3}.

Next, assume that the number of assets 
$N$ is sufficiently large (but not infinity). 
Since 
return rates $x_{i\mu}$ are independently distributed with 
mean and variance 
$0$ and $s_i$, respectively, 
and we assume that one can replace $\chi_{u\mu}$ and $\tilde{\chi}_{u\mu}$ by 
$\chi_u$ and $\tilde{\chi}_u$, respectively. 
Then, 
\bea
\f{1}{p}\sum_{\mu=1}^px_{i\mu}^2
\simeq E_X[x_{i\mu}^2]=s_i,
\eea
is approximately estimated. From this, 
\bea
\tilde{\chi}_{wi}\eq\f{p}{N}s_i\chi_u
=\a s_i\chi_u,\\
\chi_{wi}\eq\f{1}{\a s_i\chi_u},\\
\tilde{\chi}_u\eq\f{1}{N}
\sum_{i=1}^Ns_i\chi_{wi}
=\f{1}{\a\chi_u},\\
\chi_{u}\eq\f{\b}{1+\f{\b}{\a\chi_u}}
=\b\left(1-\f{1}{\a}\right),
\eea
are obtained. From these {equations}, 
the expectation of the variance of $w_i$ is 
\bea
\chi_{w}\eq\f{1}{N}\sum_{i=1}^N
\chi_{wi}=\f{\left\langle
s^{-1}
\right\rangle}{\b(\a-1)}.
\eea
This is consistent with \sref{eq66}.


\begin{thebibliography}{999}
\bibitem{Bodie}
\citebook{\citeauthorname{Z.}{Bodie}, 
\citeauthorname{A.}{Kane},
\citeauthorname{A. J.}{Marcus},}{Investments}{McGraw-Hill Education}{2014}
\bibitem{Luenberger}
\citebook{\citeauthorname{D. G.}{Luenberger},}{Investment science}{Oxford University Press}{1997}
\bibitem{Markowitz1952}
\citepaper{\citeauthorname{H.}{Markowitz},}{Portfolio selection,}{
J. Fin. {\bf 7}, 77}{1952}
\bibitem{Markowitz1959}
\citebook{\citeauthorname{H.}{Markowitz}, }{Portfolio selection: efficient diversification of 
	investments}{J. Wiley and Sons, New York.}{1959}
\bibitem{Konno}
\citepaper{\citeauthorname{H.}{Konno},
\citeauthorname{H.}{Yamazaki}, }{Mean-absolute deviation portfolio optimization model and its applications to Tokyo stock market
}{
Man. Sci. {\bf 37}, 519
}{1991}
\bibitem{Rockafellar}
\citepaper{\citeauthorname{R. T.}{Rockafellar}, 
\citeauthorname{S.}{Uryasev}, }{Optimization of conditional value-at-risk}{
J. Risk, {\bf 2},  21.}{2000}
\bibitem{Ciliberti}
\citepaper{
\citeauthorname{S.}{Ciliberti}, 
\citeauthorname{M.}{M$\acute{\rm e}$zard}, }
{Risk minimization through portfolio replication}
{
Euro. Phys. J. B, {\bf 27}, 175}{2007}
\bibitem{Pafka}
\citepaper{\citeauthorname{S.}{Pafka}, 
\citeauthorname{I.}{Kondor}, }
{Noisy covariance matrices and portfolio optimization}
{
Euro. Phys. J. B, {\bf 27}, 277}{2002}
\bibitem{Shinzato-Yasuda2015-BP}
\citepaper{\citeauthorname{T.}{Shinzato},
\citeauthorname{M.}{Yasuda},
}{Belief propagation algorithm for portfolio optimization problems}
{
PLoS One, {\bf 10}, e0134968}
{2015}
\bibitem{Shinzato2015-self-averaging}
\citepaper{\citeauthorname{T.}{Shinzato}, }
{Self-averaging property of minimal investment risk of mean-variance 
	model}
{
PLoS One, {\bf 10}, e0133846
}{2015}

\bibitem{Shinzato2008}
\citepaper{\citeauthorname{T.}{Shinzato},
\citeauthorname{Y.}{Kabashima},
}{Perceptron capacity revisited: 
	classification ability for correlated patterns}
{
J. Phys. A. {\bf 41},  324013
}
{2008}
\bibitem{Kabashima2004}
\citepaper{\citeauthorname{Y.}{Kabashima},
\citeauthorname{D.}{Saad},}
{Statistical mechanics of low-density parity-check codes}
{
J. Phys. A. {\bf 37}, R1
}
{2004}
\bibitem{Nishimori}
\citebook{\citeauthorname{H.}{Nishimori}, }
{Statistical physics of spin glasses and information 
	processing} 
{Oxford University Press}{2001}

\end{thebibliography}
\end{document}